\documentclass{article}
\usepackage{graphicx}

\begin{document}

\markboth{Zhi-Huan Luo}
{STUDY H-DIBARYON ON LATTICE QCD}

\title{H-Dibaryon from Lattice QCD with Improved Anisotropic Actions}

\author{\footnotesize
Zhi-Huan Luo\footnote{email: lozit@scau.edu.cn} \footnotesize
Mushtaq Loan\footnote{email: stsmushe@zsu.edu.cn} and Xiang-Qian
Luo\footnote{email: stslxq@zsu.edu.cn} }

\maketitle

\begin{abstract}
The six quark state(uuddss) called H dibaryon($J^P=0^+$,$S=-2$) has
been calculated to study its existence and stability. The
simulations are performed in quenched QCD on $8^3 \times 24$ and
$16^3 \times 48$ anisotropic lattices with Symanzik improved gauge
action and Clover fermion action. The gauge coupling is $\beta=2.0$
and aspect ratio $\xi=a_s/a_t=3.0$. Preliminary results indicate
that mass of H dibaryon is 2134(100)Mev on $8^3 \times 24$ lattice
and 2167(59)Mev on $16^3 \times 48$ respectively. It seems that the
radius of H dibaryon is very large and the finite size effect is
very obvious.

\end{abstract}


\section{Introduction}

In 1976, Jaffe pointed out that the quark bag model predicted the
existence of H dibaryon, which is a compound state of 6
quarks(uuddss)\cite{Jaffe:1977prl}. It is the lowest bound state in
dibaryon sector and will be a spin 0 strangeness -2 SU(3) flavor
singlet. Jaffe's original bag model suggested that $m_H=2150$Mev,
81Mev below the 2231Mev $\Lambda\Lambda$ threshold.

Since Jaffe's prediction, people tried to find out the H-Dibaryon
state both by experiment and theoretical calculation. On the
experimental side, until now, there are no enough evidences on the
existence of the H-Dibaryon. On the theoretical side, many
theoretical calculations have been maked to predict the mass of H
dibaryon. One of the most efficient ways to study this state is from
the first principle of QCD, i.e Lattice QCD. Many simulation results
suggested that H dibaryon is a bound state, but some gave contrary
conclusions
\cite{Mackenzie:1985prl,Iwasaki:1988prl,Negele:1999np,Wetzorke:2000np,Wetzorke:2003np}.

We perform the numerical simulations with refined methods, which
includes improved gauge and fermion action, smearing techniques and
tadpole improvement. To reduce computer cost and determine large
masses more accurately, we do the simulations on anisotropic lattice
for both gauge action and fermion action. This is our advantage
comparing with what the forthgoer have done.

\section{H-Dibaryon Simulation Details}

\subsection{Actions}

We generate the configurations using improved,anisotropic
action\cite{Morningstar:1997prd}:
\begin{equation}
\label{gauge action}
S_g
    =\beta\{\frac{5}{3}\frac{\Omega_{sp}}{\xi u^4_s}
    + \frac{4}{3}\frac{\xi\Omega_{tp}}{u^2_su^2_t}
    - \frac{1}{12}\frac{\Omega_{sr}}{\xi u^6_s}
    - \frac{1}{12}\frac{\xi\Omega_{str}}{u^4_su^2_t}\},
\end{equation}
where $\beta=6/g^2$, $g$ is the QCD coupling, $u_s$ and $u_t$ are
the mean-link renormalization parameters, $\xi$ is the aspect ratio
($\xi=a_s/a_t$ at the tree level in perturbation theory),and

\begin{eqnarray}
& &
 \Omega_{sp}=\sum_x\sum_{i>j}\frac{1}{3}ReTr[1-U_i(x)U_j(x+i)U^\dagger_i(x+j)U^\dagger_j(x)],
\nonumber \\
& &
\Omega_{tp}=\sum_x\sum_{i}\frac{1}{3}ReTr[1-U_i(x)U_t(x+i)U^\dagger_i(x+t)U^\dagger_t(x)],
\nonumber \\
& & \Omega_{sr}=\sum_x\sum_{i \neq
j}\frac{1}{3}ReTr[1-U_i(x)U_i(x+i)U_j(x+2i)U^\dagger_i(x+j+i)U^\dagger_i(x+j)U^\dagger_j(x)],
\nonumber \\
& &
\Omega_{str}=\sum_x\sum_{i}\frac{1}{3}ReTr[1-U_i(x)U_i(x+i)U_t(x+2i)U^\dagger_i(x+t+i)U^\dagger_i(x+t)U^\dagger_t(x)],
\nonumber
\end{eqnarray}
where $x$ labels the sites of the lattice,$i$,$j$ are spatial
indices, and $U_\mu(x)$ is the parallel transport matrix in the
gauge field from site $x$ to $x+\mu$.

For the quark action, we employ the space-time asymmetric clover
quark action on anisotropic
lattice\cite{Klassen:1999np}\cite{Okamoto:2002prd}\cite{Mei:2003ijmp}:

\begin{eqnarray}
S_f & = & \sum_x\bar{\Psi}_x\Psi_x
          \nonumber \\
    &   & -
          K_s\sum_x\sum_i[\bar{\Psi}_x(r_s-\gamma_i)U_{i,x}\Psi_{x+i}+\bar{\Psi}_x(r_s+\gamma_i)U^\dagger_{i,x-i}\Psi_{x-i}]
          \nonumber \\
    &   & -
          K_t\sum_x[\bar{\Psi}_x(1-\gamma_t)U_{t,x}\Psi_{x+t}+\bar{\Psi}_x(1+\gamma_t)U^\dagger_{t,x-t}\Psi_{x-t}]
          \nonumber \\
    &   & +
          iK_sc_s\sum_{x,i<j}\bar{\Psi}_x\sigma_{i,j}F_{ij}(x)\Psi_x
          +
          iK_sc_t\sum_{x,i}\bar{\Psi}_x\sigma_{ti}F_{ti}(x)\Psi_x,
\end{eqnarray}
where $K_{s,t}$ and $c_{s,t}$ are the spatial and temporal hopping
parameters and the clover coefficients, respectively. The hopping
parameters $K_{s,t}$ are related to the bare quark mass
$m_0=a_tm_{q0}$ through
\begin{equation}
a_tm_{q0}\equiv1/(2K_t)-3r_s/\zeta-1, \zeta=K_t/K_s.
\end{equation}

We perform the simulations using tree-level improved Symanzik action
and Clover fermion action, with gauge coupling $\beta=2.0$ and the
aspect ratio $\xi=3.0$. The lattice sizes are $8^3 \times 24$ and
$16^3 \times 48$.

\subsection{Operator and Correlation Function of H Dibaryon}
The operator of H-Dibaryon\cite{Golowich:1992prd}:
\begin{equation}
O_H(x)=3(udsuds)-3(ussudd)-3(dssduu),
\end{equation}
\begin{equation}
(abcdef)=\epsilon_{abc}\epsilon_{def}(C\gamma_5)_{\alpha\beta}(C\gamma_5)_{\gamma\delta}(C\gamma_5)_{\epsilon\phi}
         a^a_\alpha(x)b^b_\beta(x)c^c_\epsilon(x)d^d_\gamma(x)e^e_\delta(x)f^f_\phi(x),
\end{equation}
where $a,b,c,d,e,f$ are color indices and
$\alpha,\beta,\gamma,\delta,\epsilon,\phi$ are spinor indices.

And the corresponding correlation function of H Dibaryon can be
written as $G_H(\vec{x},\tau)=<O_H(\vec{x},\tau)O_H^\dagger(0)>$,
which involves terms of the structure\cite{Von:2001phd}:
\begin{equation}
(U_{11}U_{22}-U_{12}U_{21})(D_{11}D_{22}-D_{12}D_{21})(S_{11}S_{22}-S_{12}S_{21}).
\end{equation}

To decide whether the H Dibaryon is stable or not, usually we can
compare the mass of $\Lambda\Lambda$ with H-Dibaryon's. The
$\Lambda\Lambda$ operator\cite{Von:2001phd}:
\begin{equation}
O_\Lambda(x)=\epsilon_{abc}(C\gamma_5)_{\beta\gamma}[u^a_\alpha(x)d^b_\beta(x)s^c_\gamma(x)
             +d^a_\alpha(x)s^b_\beta(x)u^c_\gamma(x) -
             2s^a_\alpha(x)u^b_\beta(x)d^c_\gamma(x)].
\end{equation}

\subsection{Smearing Techniques}
To reduce the excited-state contamination, we use the smearing
techniques which can provide a better overlap with the ground
state\cite{Lacock:1995prd}. For the quark fields we use:
\begin{eqnarray}
\psi'(x,R) &=& \sum_{\mu\in V_z} (U^\dagger(x-\hat{\mu})\ldots
               U^\dagger(x-R\hat{\mu})\psi(x-R\hat{\mu})
               \nonumber \\
           & & + U(x)\ldots
               U(x+(R-1)\hat{\mu})\psi(x+R\hat{\mu})).
\end{eqnarray}

A more large plateau in the region with small errors is obtained
with smearing.

\section{Simulation Results}

Out results are shown in Tables 1, 2, 3 and 4. The $k_{H_t}$ is the
temporal heavy kappa, which corresponds to s quark; the $k_{L_t}$ is
temporal light kappa, which corresponds to u and d quarks.

\begin{table}[h]
\caption {$ma_t$ of the $\Lambda$
($8^3\times24$,$\beta=2.0$,$\xi=3.0$)} {
\begin{tabular}{|c|c|c|c|c|c|}
\hline
$k_{H_t}/k_{L_t}$ & 0.23810 & 0.23923 & 0.24039 & 0.24155 & 0.24272 \\
\hline
0.23256 & 1.3772(65) & 1.3496(67) & 1.3202(69) & 1.2885(71) & 1.2542(73) \\
\hline
0.23365 & 1.3665(66) & 1.3388(68) & 1.3093(70) & 1.2777(72) & 1.2433(74) \\
\hline
0.23474 & 1.3553(66) & 1.3276(68) & 1.2981(70) & 1.2664(72) & 1.2319(75) \\
\hline
0.23586 & 1.3438(67) & 1.3160(69) & 1.2864(71) & 1.2546(73) & 1.2201(75) \\
\hline
0.23697 & 1.3317(68) & 1.3039(69) & 1.2742(71) & 1.2424(74) & 1.2078(76) \\
\hline
0.23810 & 1.3189(69) & 1.2910(71) & 1.2613(73) & 1.2294(75) & 1.1947(77) \\
\hline
\end{tabular}
}
\end{table}

\begin{table}[h]
\caption {$ma_t$ of the $\Lambda$
($16^3\times48$,$\beta=2.0$,$\xi=3.0$)} {
\begin{tabular}{|c|c|c|c|c|c|}
\hline
$k_{H_t}/k_{L_t}$ & 0.23810 & 0.23923 & 0.24039 & 0.24155 & 0.24272 \\
\hline
0.23256 & 1.4161(32) & 1.3881(33) & 1.3583(34) & 1.3262(34) & 1.2913(35) \\
\hline
0.23365 & 1.4052(32) & 1.3772(33) & 1.3473(34) & 1.3152(34) & 1.2802(35) \\
\hline
0.23474 & 1.3940(33) & 1.3659(33) & 1.3359(34) & 1.3037(35) & 1.2688(36) \\
\hline
0.23585 & 1.3822(33) & 1.3541(34) & 1.3241(34) & 1.2918(35) & 1.2568(36) \\
\hline
0.23697 & 1.3699(33) & 1.3418(34) & 1.3117(34) & 1.2794(35) & 1.2443(36) \\
\hline
0.23810 & 1.3568(32) & 1.3286(33) & 1.2985(33) & 1.2661(34) & 1.2309(35) \\
\hline
\end{tabular}
}
\end{table}

The H-Dibaryon remain lighter than two $\Lambda$ at all combinations
of the hopping parameters, as shown in tables.

To obtain the physical masses of H and $\Lambda$, one has to
extrapolate or interpolate the $k_{L_t}$ and $k_{H_t}$ to physical
hopping parameters. Since $(m_{\pi}a)^2$ is linearly related to
$1/k$, we can determine the critical hopping parameter $k_c$ at
which $(m_{\pi}a)^2$ vanishes. We take physical $k_{ud}$ as $k_c$
because they are very close. The physical $k_s$ can be determined
from the ratio of a strangeness carrying particle to a non-strange
one, here we obtain the $k_s$ by the mass ratio of lambda and
nucleon. Our calculations suggest that $k_c=0.25256(41)$ and
$k_s=0.2413(40)$ on $8^3 \times 24$ lattice and $k_c=0.25323(18)$
and $k_s=0.2422(21)$ on $16^3 \times 48$ lattice.

\begin{table}[h]
\caption {$ma_t$ of the $H$ ($8^3\times24$,$\beta=2.0$,$\xi=3.0$)} {
\begin{tabular}{|c|c|c|c|c|c|}
\hline
$k_{H_t}/k_{L_t}$ & 0.23810 & 0.23923 & 0.24039 & 0.24155 & 0.24272 \\
\hline
0.23256 & 2.656(17) & 2.603(18) & 2.545(18) & 2.484(19) & 2.417(19) \\
\hline
0.23365 & 2.635(18) & 2.581(18) & 2.523(18) & 2.462(19) & 2.395(20) \\
\hline
0.23474 & 2.613(18) & 2.558(18) & 2.501(19) & 2.439(19) & 2.372(20) \\
\hline
0.23586 & 2.589(18) & 2.535(18) & 2.477(19) & 2.415(19) & 2.348(20) \\
\hline
0.23697 & 2.565(18) & 2.510(19) & 2.452(19) & 2.390(20) & 2.323(20) \\
\hline
0.23810 & 2.539(18) & 2.484(19) & 2.426(19) & 2.364(20) & 2.297(20) \\
\hline
\end{tabular}
}
\end{table}

\begin{table}[h]
\caption {$ma_t$ of the $H$ ($16^3\times48$,$\beta=2.0$,$\xi=3.0$)}
{
\begin{tabular}{|c|c|c|c|c|c|}
\hline
$k_{H_t}/k_{L_t}$ & 0.23810 & 0.23923 & 0.24039 & 0.24155 & 0.24272 \\
\hline
0.23256 & 2.7641(87) & 2.7093(90) & 2.6508(93) & 2.5882(97) & 2.5203(102) \\
\hline
0.23365 & 2.7422(88) & 2.6872(91) & 2.6287(94) & 2.5660(98) & 2.4980(103) \\
\hline
0.23474 & 2.7194(89) & 2.6643(92) & 2.6057(95) & 2.5429(99) & 2.47489(105) \\
\hline
0.23585 & 2.6956(90) & 2.6405(93) & 2.5818(97) & 2.5188(101) & 2.4507(106) \\
\hline
0.23697 & 2.6708(91) & 2.6155(94) & 2.5567(98) & 2.4937(102) & 2.4254(108) \\
\hline
0.23810 & 2.6407(86) & 2.5854(88) & 2.5265(92) & 2.4634(96) & 2.3951(101) \\
\hline
\end{tabular}
}
\end{table}

In fig.1, we performed a linear fit to extrapolate the $m_H$ and
$m_{\Lambda}$ to the physical $k_s$, and obtained the H's mass
$m_H=2134(100)$Mev, which is lower than two $\Lambda's$. The
difference in mass is $m_H-2m_{\Lambda}=-97(100)$Mev. That means the
H-Dibaryon tends to be a bound state but actually we can't make this
conclusion because the error is larger than the mass difference on
$8^3\times24$ lattice. In fig.2, on larger lattice($16^3 \times
48$), $m_H=2167(59)$Mev and $m_H-2m_{\Lambda}=-64(59)$Mev, indicates
that the energey of H dibaryon tends to below the $\Lambda\Lambda$
threshold.

\includegraphics[width=4.0in]{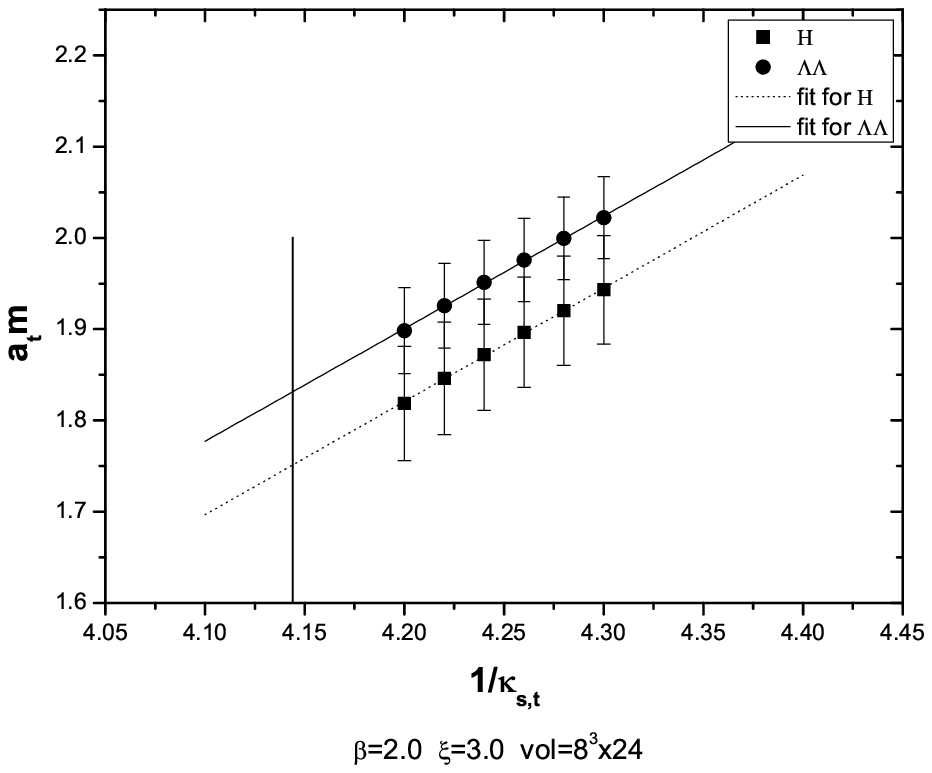}
\\
\includegraphics[width=4.0in]{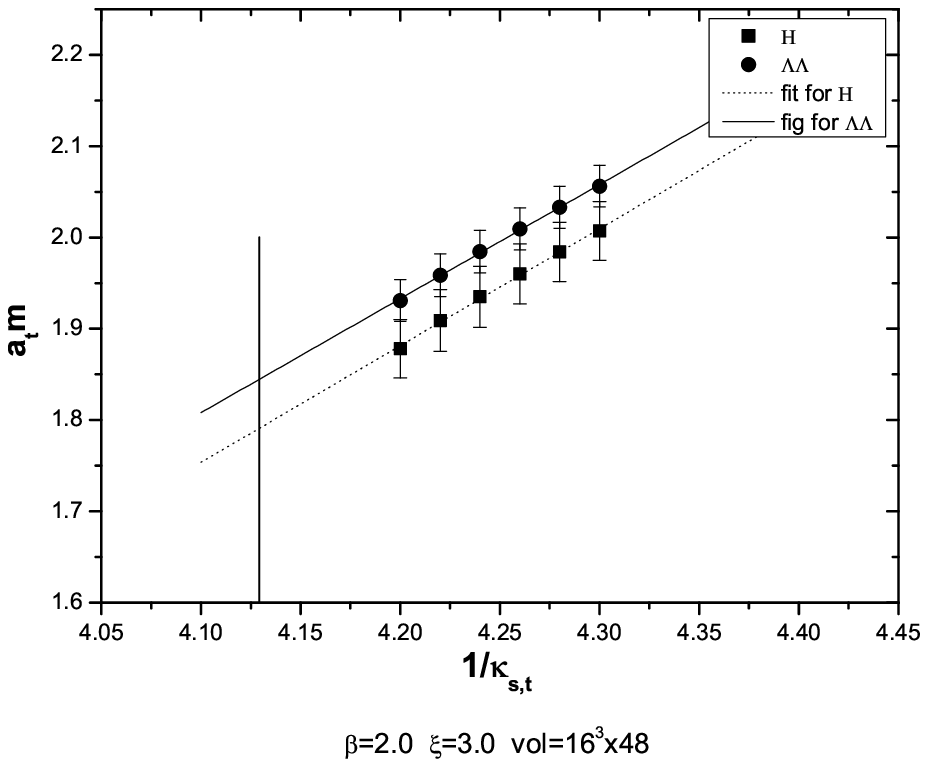}

\section{Conclusion and Future Plans}

We have presented the results of the lattice investigation on the H
dibaryon state employing anisotropic improved gauge and anisotropic
clover fermion actions. The advantage of using anisotropic QCD
actions is to get get a better signal so as to obtain a large
plateau on small lattice size which can save us much more computer
cost. In the mean time, the simulation results are more accurately.

Our results indicate that, both on the $8^3 \times 24$ and $16^3
\times 24$ anisotropic lattice, the masses of H dibaryon are less
than that of two $\Lambda$s. It seems that the H dibaryon do exist
as a bound state.

The masses of H Dibaryon on two different lattices are not so close
and we believe that the finite size effect of H dibaryon should be
taken into account. We plan to calculate the H dibaryon on larger
lattice size to further study the finite size effect. We also intend
to calculate other six quarks states, such as possible
proton-antiproton, deuteron, and so on. It should be very
interesting.

\section{Acknowledgement}

The code is based on MILC's Code. The authors are grateful to
Carleton DeTar and Ines Wetzorke for fruitful discussions, and to
ICMSEC(The Institute of Computational Mathematics and
Scientific/Engineering Computing of Chinese Academy of Sciences) for
providing the LSSC2 cluster on which some of our simulations were
performed.

\end{document}